\documentclass[12pt]{article}

\usepackage{amsmath,amssymb}
\usepackage{cite}

\usepackage[unicode,linktocpage=true,plainpages=false,pdfpagelabels=false]{hyperref}

\newcommand{\dd}{\partial}
\newcommand{\de}{\delta}
\newcommand{\m}{\mu}
\newcommand{\n}{\nu}
\newcommand{\ls}{\left(}
\newcommand{\rs}{\right)}
\newcommand{\la}{\lambda}
\newcommand{\ka}{\varkappa}
\newcommand{\ga}{\gamma}
\newcommand{\ta}{\tau}
\newcommand{\al}{\alpha}
\newcommand{\be}{\beta}
\newcommand{\ep}{\varepsilon}
\newcommand{\ps}{\psi}

\newcommand{\disn}[2]{$$\displaylines{\refstepcounter{equation}%
            \label{#1}\hskip 1em minus 1em #2\hfilneg}$$}
\newcommand{\nom}{\hfil\hskip 1em minus 1em (\theequation)}
\newcommand{\no}{\hfil \hskip 1em minus 1em\phantom{(\theequation)}%
            \hfilneg\cr\hfilneg\hskip 1em minus 1em\hfil}
\newcommand{\ns}{\hfill\cr\hfill}

\begin{document}

\title{Forms of action for perfect fluid in General Relativity\\ and mimetic gravity}
\author{S.~A.~Paston\thanks{E-mail: s.paston@spbu.ru}\\
{\it Saint Petersburg State University, Saint Petersburg, Russia}
}
\date{\vskip 15mm}
\maketitle

\begin{abstract}
We consider various forms of action allowing us to describe a perfect fluid in the framework of General Relativity.
First we consider a potential motion without pressure.
Starting from an action in terms of the current density vector built  similarly to the action of  a single particle,
we build a series of equivalent actions through various changes of variables.
The results are then generalized to the cases of motion with vorticity and the presence of pressure.
The obtained forms of action are compared to the previously known variants.
We discuss the possibilities to use the obtained results to formulate the theories containing a description of a perfect fluid.
A new form of action is proposed for mimetic gravity, which allows a motion with vorticity for mimetic matter as a perfect fluid.
\end{abstract}

\newpage

\section{Introduction}\label{rg-vved}
In recent years, the idea of mimetic gravity proposed in \cite{mukhanov}, in which the conformal degrees of freedom of gravity are isolated by introducing a parametrization of the physical metric under the form
\disn{v1}{
g_{\m\n}=\tilde g_{\m\n}\tilde g^{\ga\de}(\dd_\ga\la)(\dd_\de\la),
\nom}
has become quite popular.
This approach uses a conventional form of the Einstein-Hilbert gravitational action for the physical metric $g_{\m\n}$,
but for the independent variables, an auxiliary metric $\tilde g_{\m\n}$ and a scalar field $\la$ are taken.
As a result, the equations of motion take the form of the Einstein equations with an additional contribution to the energy momentum tensor (EMT) from the mimetic matter.
The mimetic matter has in this case the properties of a perfect fluid with a \emph{zero pressure} in a \emph{potential motion},
i.~e. its 4-speed has the form of
\disn{v2}{
u_\m=\dd_\m\la,\qquad u^\m u^\n g_{\m\n}=1
\nom}
(the signature $+---$ is used).
In such a way a "{}`dark matter' without dark matter, which is imitated by extra scalar degree of freedom of the gravitational field"{} \cite{mukhanov} arises in the theory.
After some modifications
(we briefly discuss one of the directions in Section~\ref{rg-mimet})
of mimetic gravity within the framework of such an approach, it becomes possible to explain some of the existing problems of modern cosmology;
see the review \cite{mimetic-review17} and references therein for the current state of this subject.

It has been noted in \cite{Golovnev201439} that the equations of motion of mimetic gravity can be obtained from the action containing the Lagrange multiplier $n$
\disn{v3}{
S=-\frac{1}{2\ka}\int\! d^4x\sqrt{-g}\,R-
\int\! d^4x\sqrt{-g}\, \Bigl( 1-g^{\m\n}(\dd_\m\la)(\dd_\n\la)\Bigr)n,
\nom}
where $g$ is the determinant of the $g_{\m\n}$ metric, $R$ is the scalar curvature and $\ka$ is the gravitation constant.
The "mimetic"{} idea is no longer used in this action,
which is a usual General Relativity action with some matter described by two scalar fields $n$ and $\la$,
and this matter appears to be a \emph{potentially} moving perfect fluid \emph{without pressure}. Following the authors of the overview \cite{mimetic-review17} we note that the action \eqref{v3} is a particular case of the action suggested in \cite{lim1003.5751}, for which the motion is also potential, while the pressure is not null.
A somewhat redundant (see the discussion below, after \eqref{vi5}) generalization of \eqref{v3} to the case of nonpotential (with vorticity) motion in accordance with the idea \cite{lin1963} is given in \cite{brown93}.

We analyze now  in what other form (probably more suitable for applications), we can write the action for a perfect fluid, while rejecting, if possible, the limitations of the potentiality of motion and of the absence of pressure.
Solving this question can be useful for the study of cosmological problems for further development of mimetic gravity -- as a result of taking into account the motion with vorticity of the mimetic matter
(see the mimetic gravity action in Section~\ref{rg-mimet})
as well as the presence of pressure
(the latter was already used in \cite{lim1003.5751} to describe the dark energy),
as well as in other cases.
For example, in the phenomenological study of particle creation induced by strong gravitational fields \cite{berezin17} a hydrodynamical description of the particle motion is used,
where the particle number conservation law is modified.
The possibility to write a hydrodynamical action in different ways can be of use for the development of this approach.

Several variants of the action of perfect fluid were proposed and investigated in a number of works
\cite{lin1963,seliger1968,schutz1970,ray1972,berezin1987,brown93,Fukagawa10,dieztejedor13,ariki16,minguzzi16}.
Below, as far as the results will be obtained, we will specify their relation to the results of these works.
In a part of them the Lagrange multipliers were used for the construction of the action in the same way as in the above-mentioned works \cite{lim1003.5751,Golovnev201439}.
In the present paper we will be based on the idea of constructing the action for a perfect fluid  using the current vector $j^\m$ as an initial independent variable.
Then, substituting, removing and adding some variables in this action, we will obtain several equivalent forms of the action.
Some of them are related to previously known variants, while some others are new.
In Section~\ref{rg-prosto} we restrict ourselves to the case of the potential motion of a perfect fluid without pressure;
in Section~\ref{rg-vihr} we generalize the results to the case of motion with vorticity.
In Section~\ref{rg-mimet} a new variant of the mimetic gravity action for which the mimetic matter is a perfect fluid with vorticity is suggested,
and in Section~\ref{rg-davl} we generalize the results of Sections~\ref{rg-prosto},\ref{rg-vihr} to the case of a pressure.

In the Conclusion we consider again the possible use of the obtained results.

\section{Potential motioned perfect fluid without pressure}\label{rg-prosto}
Following a direct analogy with the action of a point unit mass relativistic particle
\disn{p1}{
S=-\int\! d\ta\, \sqrt{\dot x^\m(\ta)\dot x^\n(\ta)g_{\m\n}(x(\ta))}
\nom}
we write a perfect fluid action with no pressure under the form
\disn{p2}{
S_1=-\int\! d^4 x\, \sqrt{-g}\ls \sqrt{j^\m j^\n g_{\m\n}}-j^\m \dd_\m\la\rs,
\nom}
where $j^\m$ is the current density vector of the liquid and $\la$ is a scalar that appears to be the Lagrange multiplier,
the addition of which ensures the continuity equation for the current.
Varying with respect to $j^\m$ and $\la$, one can easily obtain the equations of motion, which with the notations
\disn{p3}{
n\equiv \sqrt{j^\m j^\n g_{\m\n}},\qquad
u^\m\equiv \frac{j^\m}{\sqrt{j^\m j^\n g_{\m\n}}}
\nom}
(these values are interpreted as a particle number density and the unit four-velocity)
take the form
\disn{p4}{
D_\m (n u^\m)=0,\qquad
u_\m=\dd_\m\la,
\nom}
where $D_\m$ is the covariant derivative.
It follows from the last equation that
\disn{p5}{
D_\m u_\n-D_\n u_\m=0\quad\Rightarrow\quad
u^\m D_\m u_\n=0,
\nom}
i.~e. the fluid moves along the geodesics, but this is necessarily a \emph{potential} motion.

We easily note that the EMT corresponding to \eqref{p2}
\disn{p5.1}{
T^{\m\n}=n u^\m u^\n
\nom}
corresponds to a perfect fluid without pressure.
Note that the action \eqref{p2} also can be rewritten,
assuming that the independent variable is not the vector $j^\m$, but the vector density $J^\m=\sqrt{-g}j^\m$ which is related to it.
In this case, neither the equations of motion not the expression for the EMT changes.
The generalization of the action \eqref{p2} to the case of nonpotential motion is written in \cite{brown93} just in the terms of  $J^\m $.
This generalization is based on the idea of the work \cite{lin1963}, see details in the Section~\ref{rg-vihr}.

We will find now an alternate expression, containing no radical, for the action of a perfect fluid without pressure.
We use again the analogy with the action for a particle, which can (see, for example~\cite{GreenSchwarzWitten})  be rewritten under an equivalent form with no radical:
\disn{p5.2}{
S=-\frac{1}{2}\int\! d\ta\,\ls \frac{1}{e(\ta)}\dot x^\m(\ta)\dot x^\n(\ta)g_{\m\n}(x(\ta))+e(\ta)\rs.
\nom}
Similarly we obtain
\disn{p6}{
S_2=-\int\! d^4 x\, \sqrt{-g}\ls  \frac{1}{2q}j^\m j^\n g_{\m\n}+\frac{q}{2}-j^\m \dd_\m\la\rs.
\nom}
It is easy to verify that varying with respect to $q$ gives the equation $q=\sqrt{j^\m j^\n g_{\m\n}}$,
while varying with respect to $j^\m$ and $\la$ gives again the equations \eqref{p4},
i.~e. the theories \eqref{p2} and \eqref{p6} are classically equivalent.
The EMT for the action \eqref{p6} is given again by \eqref{p5.1}.

We easily note that the action \eqref{p6} is nonpolynomial in one of its scalar variables (in $q$),
which can cause problems in some cases.
It can be easily made polynomial by substituting $q=a^2$, $j^\m=a\tilde j^\m$, resulting in
\disn{p7}{
S_3=-\int\! d^4 x\, \sqrt{-g}\ls\frac{1}{2}\tilde j^\m \tilde j^\n g_{\m\n}+\frac{a^2}{2}-a\tilde j^\m \dd_\m\la\rs.
\nom}
The motion equations for this action are
\disn{p8}{
D_\m(a \tilde j^\m)=0,\qquad
\tilde j_\m=a\dd_\m\la,\qquad
a=\tilde j^\m \dd_\m\la,
\nom}
reproducing again the equations \eqref{p4} and resulting in the expression for the EMT \eqref{p5.1},
so \eqref{p7} is a correct action for a potentially moving perfect fluid without pressure.

This kind of action has an interesting property: it depends quadratically both on the vector variable $\tilde j^\m$ and on the scalar one $a$.
This gives two different ways to proceed to a new theory,
equivalent to \eqref{p7} not only in the classical sense, but, to some extent, even in the quantum sense,
meaning the possibility of taking a Gaussian functional integral (neglecting the  quantum determinant).	
The first way is to substitute in the action the expression \eqref{p8} for $a$ (in the quantum theory to take a functional integral over $a$), resulting in the action
\disn{p9}{
S_4=-\frac{1}{2}\int\! d^4 x\, \sqrt{-g}\,\Bigl( g_{\m\n}-(\dd_\m\la)(\dd_\n\la)\Bigr)\tilde j^\m \tilde j^\n.
\nom}
The motion equations for this action are
\disn{p10}{
\Bigl( g_{\m\n}-(\dd_\m\la)(\dd_\n\la)\Bigr)\tilde j^\n=0,\qquad
D_\m\ls \tilde j^\m \tilde j^\n\dd_\n\la\rs=0.
\nom}
It follows from the first equation that
$g^{\m\n}(\dd_\m\la)(\dd_\n\la)=1$,
and that
$\tilde j_\m\sim\dd_\m\la$,
giving again the initial motion equations \eqref{p4} and the EMT \eqref{p5.1}, assuming that
\disn{p11}{
n=g_{\m\n}\tilde j^\m\tilde j^\n,\qquad
u^\m\equiv \frac{\tilde j^\m}{\sqrt{\tilde j^\m \tilde j^\n g_{\m\n}}}.
\nom}

Consider now the second of the above two methods of simplifying the action \eqref{p7} -- substitute in it the expression \eqref{p8} for $\tilde j^\m$
(take a functional integral over $\tilde j^\m$ in the quantum theory),
resulting in the action
\disn{p12}{
S_5=-\frac{1}{2}\int\! d^4 x\, \sqrt{-g}\,\Bigl( 1-g^{\m\n}(\dd_\m\la)(\dd_\n\la)\Bigr)n,
\nom}
where the value $a^2$ was replaced by $n$.
Comparing this expression to \eqref{v3} we easily note that the obtained form of the action differs only by a numerical factor from the one proposed in \cite{Golovnev201439} as equivalent to the action of mimetic gravity.
As already stated in the Introduction, various generalizations of the action \eqref{p12} were discussed in
\cite{brown93} and \cite{lim1003.5751}.
Only two scalar independent variables remained in this action.  Varying with respect to them, we easily found the motion equations
\disn{p13}{
g^{\m\n}(\dd_\m\la)(\dd_\n\la)=1,\qquad
D_\m \ls n g^{\m\n}(\dd_\n\la)\rs=0,
\nom}
which coincide with \eqref{p4} if we denote $u_\m\equiv\dd_\m\la$, when the EMT coincides again with \eqref{p5.1}.

Note an alternative possibility to make the action \eqref{p6} polynomial, besides that described above.
Polynomiality can be achieved by introducing one more independent scalar field $n$:
\disn{p14}{
S_6=-\int\! d^4 x\, \sqrt{-g}\,\Big( n-j^\m \dd_\m\la+\ga\ls j^\m j^\n g_{\m\n}-n^2\rs\Bigr),
\nom}
where $\ga$ is the Lagrange multiplier related to the field $q$ from \eqref{p6} by $\ga=1/(2q)$.
The action \eqref{p6} can be obtained from \eqref{p14} by excluding the variable $n$, which is expressed from the equations of motion and is substituted into the action.
Since $n$ is quadratic in the action, the transition from the theory \eqref{p14} to \eqref{p6} can also be understood in the quantum sense, in the form of taking a functional Gaussian integral.
The motion equations for \eqref{p14} are
\disn{p15}{
 j^\m j^\n g_{\m\n}=n^2,\qquad
2\ga n=1,\qquad
 2\ga j_\m=\dd_\m\la,\qquad
 D_\m j^\m=0
\nom}
and we easily note that they give the motion equations \eqref{p4} and EMT \eqref{p5.1}.
The action \eqref{p14} is a special case of the action used in \cite{dieztejedor13}, restricted to the case of an isentropic pressureless fluid.

Thus all the six forms of the action \eqref{p2}, \eqref{p6}, \eqref{p7}, \eqref{p9}, \eqref{p12}, \eqref{p14}
describe the same system: a potentially moving perfect fluid without pressure,
and they can be used as appropriate in particular tasks.
In the next sections the obtained results will be generalized to the cases of a nonpotential motion and the presence of pressure.

\section{Motion with vorticity}\label{rg-vihr}
All the forms of action from the previous section describe the motion of a perfect fluid with no vorticity, due to the fact that in \eqref{p4} the speed $u_\m$ has a gradient form.
To eliminate this limitation we will modify these actions following the idea of \cite{seliger1968}.
We write the modification of the action \eqref{p2} as
\disn{vi2}{
S_1=-\int\! d^4 x\, \sqrt{-g}\ls \sqrt{j^\m j^\n g_{\m\n}}-j^\m \ls\dd_\m\la+\al\dd_\m\be\rs\rs,
\nom}
thus adding to the theory two independent scalar variables $\al$ and $\be$.
The equations of motion corresponding to this action, with the notations \eqref{p3},
are
\disn{vi4}{
D_\m (n u^\m)=0,\qquad
u_\m=\dd_\m\la+\al\dd_\m\be,\qquad
u^\m\dd_\m\al=0,\qquad
u^\m\dd_\m\be=0.
\nom}
Using the last three equations we obtain
\disn{vi5}{
D_\m u_\n-D_\n u_\m=(\dd_\m\al)(\dd_\n\be)-(\dd_\n\al)(\dd_\m\be)\quad\Rightarrow\quad
u^\m D_\m u_\n=0,
\nom}
i.~e. the liquid moves as before along the geodesics. However, this motion is not potential now, and it can be quite arbitrary according Clebsch representation \cite{Clebsch1859} for the speed, see details in \cite{seliger1968,Fukagawa10}.
We easily verify that, taking into account the equations of motion \eqref{vi4}, corresponding to the action \eqref{vi2} the EMT expression is reduced to \eqref{p5.1} as in the previous section.

Note that the idea of \cite{seliger1968} used in \eqref{vi2}
is an improvement of earlier ideas of the work \cite{lin1963}, where instead of two fields $\al,\be$ two triples of such fields were introduced, one of these triples being understood as the "number"{} of a particle in the stream.
However, this approach turned out to be redundant, as noticed in \cite{seliger1968} as a result of the use of the Clebsch representation.
The idea of using only two fields $\al,\be$ was also suggested later in \cite{ray1972}.
A variant of the action \eqref{vi2} using not two fields $\al,\be$, but two triples of such fields, was suggested (in terms of $J^\m=\sqrt{-g}j^\m$, but not $j^\m$) in \cite{brown93}.

We write here the modifications of the remaining five actions \eqref{p6}, \eqref{p7}, \eqref{p9}, \eqref{p12}, \eqref{p14} from the previous section, which allow us
to describe the vortical motion:
\disn{vi6}{
S_2=-\int\! d^4 x\, \sqrt{-g}\ls  \frac{1}{2q}j^\m j^\n g_{\m\n}+\frac{q}{2}-j^\m \ls\dd_\m\la+\al\dd_\m\be\rs\rs,
\nom}
\disn{vi7}{
S_3=-\int\! d^4 x\, \sqrt{-g}\ls\frac{1}{2}\tilde j^\m \tilde j^\n g_{\m\n}+\frac{a^2}{2}-a\tilde j^\m \ls\dd_\m\la+\al\dd_\m\be\rs\rs,
\nom}
\disn{vi9}{
S_4=-\frac{1}{2}\int\! d^4 x\, \sqrt{-g}\,\Bigl( g_{\m\n}-\ls\dd_\m\la+\al\dd_\m\be\rs\ls\dd_\n\la+\al\dd_\n\be\rs\Bigr)\tilde j^\m \tilde j^\n,
\nom}
\disn{vi12}{
S_5=-\frac{1}{2}\int\! d^4 x\, \sqrt{-g}\,\Bigl( 1-g^{\m\n}\ls\dd_\m\la+\al\dd_\m\be\rs\ls\dd_\n\la+\al\dd_\n\be\rs\Bigr)n,
\nom}
\disn{vi14}{
S_6=-\int\! d^4 x\, \sqrt{-g}\,\Big( n-j^\m \ls\dd_\m\la+\al\dd_\m\be\rs+\ga\ls j^\m j^\n g_{\m\n}-n^2\rs\Bigr).
\nom}
For brevity, we will not discuss the corresponding equations of motion and the expressions for the EMT,
because it can easily done, following the logic described in the previous section.
For each action, one can see that the equations of motion and the expressions for the EMT are reduced to the formulas \eqref{vi4} and \eqref{p5.1} taking into account the notations \eqref{p3}.

\section{Mimetic matter with vorticity}\label{rg-mimet}
As mentioned in the Introduction, after the appearance of the mimetic matter idea in the work \cite{mukhanov}
various variants of generalization of the model were considered in subsequent works.
In particular, these generalizations can remove the restriction of the potentiality of the mimetic matter motion.
One of the most interesting variants is probably the addition of terms containing the potential of field $\la$ and(or) its higher derivatives to the action \eqref{p12}.
As a result the action is reduced to the sum of the Einstein-Hilbert term and of the expression
\disn{vi14.1}{
-\frac{1}{2}\int\! d^4 x\, \sqrt{-g}\,\biggl(\Bigl( 1-g^{\m\n}(\dd_\m\la)(\dd_\n\la)\Bigr)n-
V(\la)+\frac{1}{2}\ga(\la)\ls\Box\la\rs^2\biggr),
\nom}
where $\Box\equiv g^{\m\n}D_\m D_\n$ and $V(\la)$,$\ga(\la)$ are some given functions.
Within the context of the mimetic gravity approach the influence of the term $V(\la)$ was studied in \cite{mukhanov2014}
(note that such a structure of this action, as well as the action \eqref{p12}, appears to be a particular case of the general expression,
already considered in the earlier work mentioned above \cite{lim1003.5751}).
Adding a term with higher derivatives of the field $\la$ with a constant coefficient $\ga$ was also suggested in \cite{mukhanov2014},
while in the work \cite{vikman2015} this coefficient was made dependent on $\la$, and the potential $V(\la)$ was not introduced.

If we use in \eqref{vi14.1} only the potential $V(\la)$, but not the term with higher derivatives,
then the emerging theory (as well as a more general theory considered in \cite{lim1003.5751}) describes a perfect fluid with pressure, but with no vorticity.
With the account of higher derivatives the \emph{vorticity} emerges, but the theory describes in this case the motion of an \emph{imperfect fluid} \cite{vikman2015}.
The mimetic gravity with the mimetic matter as an imperfect fluid was studied, for example, in a recent paper \cite{kobayashi2017},
which suggested passing to the mimetic scalar-tensor theory (studied also in \cite{arroja2015}) to solve the instability problem.

However, if we use the forms of the action \eqref{vi2}, \eqref{vi6}-\eqref{vi14} obtained in the previous Section,
we can try to obtain a mimetic gravity with the mimetic matter being a \emph{perfect fluid}, but with \emph{vorticity}.
It can be done directly based on the action \eqref{vi12} containing the scalar fields only, if we pass from it to an equivalent action
\disn{vi15}{
S=-\frac{1}{2\ka}\int\! d^4x\sqrt{-g}\,R(g),
\nom}
where the physical metric $g_{\m\n}$ is expressed using the auxiliary metric $\tilde g_{\m\n}$ and the scalar fields $\la,\al,\be$ by a formula generalizing \eqref{v1}
\disn{vi16}{
g_{\m\n}=\tilde g_{\m\n}\tilde g^{\ga\de}\ls\dd_\ga\la+\al\dd_\ga\be\rs\ls\dd_\de\la+\al\dd_\de\be\rs.
\nom}
One can easily verify that with the notations
\disn{vi17}{
n\equiv g_{\m\n}\ls \frac{1}{\ka}G^{\m\n}-T^{\m\n}\rs,\qquad
u_\m\equiv\dd_\m\la+\al\dd_\m\be
\nom}
the motion equations for this action are
\disn{vi18}{
G^{\m\n}=\ka \ls T^{\m\n}+n u^\m u^\n\rs,\qquad
D_\m (n u^\m)=0,\no
g^{\m\n}u_\m u_\n=1,\qquad
u^\m\dd_\m\al=0,\qquad
u^\m\dd_\m\be=0.
\nom}
With the account of \eqref{vi5} it follows from them that the mimetic matter moves along the geodesics, but this motion is not necessarily potential.

The mimetic matter action \eqref{vi15},\eqref{vi16} can be directly related to the perfect fluid action \eqref{vi12}, following the idea suggested in \cite{Chaichian:2014qba}\footnote{
The author is grateful to the Phys.~Rev.~D reviewer having indicated this possibility.}.
To do this we introduce the notation
\disn{vi18.1}{
\Phi^2=\tilde g^{\ga\de}\ls\dd_\ga\la+\al\dd_\ga\be\rs\ls\dd_\de\la+\al\dd_\de\be\rs,
\nom}
and we obtain $g_{\m\n}=\tilde g_{\m\n}\Phi^2$.
It is well known that with such Weyl transformations of the auxiliary metric $\tilde g_{\m\n}$ the Einstein-Hilbert action \eqref{vi15} transforms into the action
\disn{vi18.2}{
S=-\frac{1}{2\ka}\int\! d^4x\sqrt{-\tilde g}\,\Bigl( R(\tilde g)\Phi^2+6\tilde g^{\m\n}(\dd_\m\Phi)(\dd_\n\Phi)\Bigr)
\nom}
up to the surface terms.
Introducing the additional Lagrange multiplier $n$, we can proceed to an equivalent action
\disn{vi18.3}{
S=-\frac{1}{2\ka}\int\! d^4x\sqrt{-\tilde g}\,\biggl( R(\tilde g)\Phi^2+6\tilde g^{\m\n}(\dd_\m\Phi)(\dd_\n\Phi)+\ns+
\ka\Bigl(\Phi^2-\tilde g^{\m\n}\ls\dd_\m\la+\al\dd_\m\be\rs\ls\dd_\n\la+\al\dd_\n\be\rs\Bigr)n
\biggr),
\nom}
where the field $\Phi$ is no longer expressed \emph{a priori} by \eqref{vi18.1}, but it is arbitrary, satisfying the equation \eqref{vi18.1} only due to the motion equation arising from varying with respect to $n$.
We easily note that, since the action \eqref{vi15},\eqref{vi16} is gauge invariant with respect to Weyl transformations of the auxiliary metric $\tilde g_{\m\n}$,
the actions \eqref{vi18.2} and \eqref{vi18.3} also have the same invariance (up to the account of the surface terms).
In the latter one, if the metric is scaled, the scalar fields  $\Phi$ and $n$ are also scaled correspondingly.
We can choose the condition $\Phi=1$ as the gauge fixing for this gauge invariance.
As a result, the action \eqref{vi18.3} transforms into the sum of the Einstein-Hilbert action  and the perfect fluid action \eqref{vi12}, where $\tilde g_{\m\n}$ is used as a metric.

\section{Case of the presence of pressure}\label{rg-davl}
Now, we build the modifications of the forms of action obtained in section~\ref{rg-vihr}, allowing us to consider the pressure of the fluid.
In this case the motion of the liquid is quite arbitrary because of the presence of scalar fields $\al,\be$,
but if they are removed, the motion will be subject to the constraints corresponding to a potential motion in the pressureless case.

Consider first the modification of the action \eqref{vi2} under the form
\disn{di2}{
S_1=-\int d^4 x \sqrt{-g}\ls \ep\ls\sqrt{j^\m j^\n g_{\m\n}}\rs-j^\m\ls \dd_\m\la+\al\dd_\m\be\rs\rs,
\nom}
where $\ep(n)$ is a given function (recall that the value $n$ is defined by \eqref{p3} and interpreted as a particle number density).
The motion equations in such a theory with the notations \eqref{p3} can be written in the form generalizing \eqref{vi4} as
\disn{d3}{
D_\m (n u^\m)=0,\qquad
\ep'(n)u_\m=\dd_\m\la+\al\dd_\m\be,\qquad
u^\m\dd_\m\al=0,\qquad
u^\m\dd_\m\be=0,
\nom}
where the  prime means the differentiation with respect to the argument of the function.
The expression for the EMT reads
\disn{d4}{
T^{\m\n}=-\frac{2}{\sqrt{-g}}\frac{\de S_1}{\de g_{\m\n}}=
\ep'(n)n u^\m u^\n+g^{\m\n}\Bigl( \ep(n)-n u^\m\ls \dd_\m\la+\al\dd_\m\be\rs\Bigr)=\ns=
\ep'(n)n u^\m u^\n+g^{\m\n}\ls \ep(n)-n \ep'(n)\rs,
\nom}
where the motion equations \eqref{d3} are used. Comparing it to the standard form of the EMT of a perfect fluid $T^{\m\n}=(\ep+p)u^\m u^\n-p g^{\m\n}$,
one can conclude that the action \eqref{di2} describes a perfect fluid,
if the value $\ep(n)$ is identified with the energy density and the pressure $p$ is defined by the formula
\disn{d5}{
p=n \ep'(n) - \ep(n).
\nom}
In this case, according to the second formula \eqref{d3} at a given $n$, the speed vector $u_\m$ is parametrized by three scalars $\la,\al,\be$, so the motion of the fluid appears to be arbitrary enough; see the text after \eqref{vi5}. The choice of the dependence $\ep(n)$ (the energy density on the particle  number density) in accordance with \eqref{d5} defines the equation of state $p(\ep)$.
At a given equation of state, one can easily obtain from \eqref{d5} and then solve a differential equation for finding the function $n(\ep)$ inverse to $\ep(n)$:
\disn{d6}{
p(\ep)=n(\ep)\ls\frac{dn}{d\ep}\rs^{-1} - \ep\quad\Rightarrow\quad
n(\ep)=n_0\exp\int\frac{d\ep}{\ep+p(\ep)}.
\nom}
For example, for the equation of state $p(\ep)=w\ep$, with $w$ constant, one easily obtains
\disn{d7}{
n(\ep)=\ls\frac{\ep}{C}\rs^\frac{1}{1+w}\quad\Rightarrow\quad
\ep(n)=C\, n^{1+w},
\nom}
where $C$ is the dimensional integration constant determining the energy of one particle.
For $w=0$ we naturally return to the theory \eqref{vi2} without pressure.
It is interesting to note that the case $w=-1$ corresponding to the dark energy
appears to be special from the point of view of this consideration.

An action similar to \eqref{di2} (in terms of not $j^\m$, but $J^\m=\sqrt{-g}j^\m$) was considered in \cite{brown93},
however, instead of two scalar fields $\al,\be$, two triples of such fields were redundantly introduced.
The nuances of these two approaches have already been considered in Section \ref{rg-vihr}, see the next paragraph after \eqref{vi5}.
In \cite{brown93} the same expressions were obtained for the EMT \eqref{d4}, and therefore for the pressure \eqref{d5}.
Similar considerations were also made in \cite{minguzzi16}, but using a modification of the variational principle instead of the Lagrange multipliers.
It resulted in an expression for EMT other than \eqref{d4}, and the interpretation of the results turned out to be different.

Consider now the alternative variants of the action \eqref{di2} built in the same way as in the Sections~\ref{rg-prosto},\ref{rg-vihr}.
The generalization of the actions \eqref{p6},\eqref{vi6} is
\disn{di6}{
S_2=-\int\! d^4 x\, \sqrt{-g}\ls  \frac{1}{2q}j^\m j^\n g_{\m\n}+\frac{1}{2}f(q)-j^\m \ls\dd_\m\la+\al\dd_\m\be\rs\rs.
\nom}
Varying it with respect to $q$ gives the equation
\disn{d8}{
f'(q)=\frac{n^2}{q^2},
\nom}
where the notation \eqref{p3} is used again.
We can express $q$ from \eqref{d8} and substitute it into the action \eqref{di6}. The obtained action will then coincide with \eqref{di2} if the following equation is satisfied:
\disn{d9}{
\frac{n^2}{2q}+\frac{f(q)}{2}=\ep(n)\quad\Rightarrow\quad
qf'(q)+f(q)=2\ep(n).
\nom}
Note that if we differentiate the first equation \eqref{d9} with respect to $n$ using \eqref{d8} we obtain
\disn{d9.1}{
\ep'(n)=\frac{n}{q}.
\nom}
Substituting this equality and \eqref{d8} in \eqref{d5} we can write the latter as
\disn{d10}{
qf'(q)=\ep+p,
\nom}
where the pressure $p$ will be assumed as a given function of the energy density $\ep$.
From \eqref{d9} and \eqref{d10} we can conclude that $f(q)=\ep-p$. Denote $\xi\equiv\ep-p$ and assume that $\ep$ and $p$ are some functions of $\xi$.
Then \eqref{d10} can be rewritten as an easily solvable equation for the function $q(\xi)$ inverse to $f(q)$:
\disn{d11}{
\frac{dq}{d\xi}=\frac{q}{\ep(\xi)+p(\xi)}\quad\Rightarrow\quad
q=q_0\exp\int\frac{d\xi}{\ep(\xi)+p(\xi)}.
\nom}
Inverting this dependence allows us to find the function $f(q)$ in the action \eqref{di6} for any given equation of state $p(\ep)$.
For the equation of state $p(\ep)=w\ep$ one easily obtains
\disn{d12}{
f(q)=Cq^{\frac{1+w}{1-w}},
\nom}
where $C$ is the dimensional integration constant determining again the energy of one particle.
For $w=0$ we return again to the theory \eqref{vi6} without pressure.
Note that for the action \eqref{di6} not only does the dark energy case $w=-1$ turn out to be special, but so also does the case $w=1$, for which $f(q)=0$.
It is easy to verify, using the relations \eqref{d9},\eqref{d9.1}, that the equations of the theory \eqref{di6} coincide with \eqref{d3} and
the EMT coincides with \eqref{d4}.

Next we write the generalization of the actions \eqref{p7},\eqref{vi7}.  To do this we substitute in \eqref{di6} $q=a^2$, $j^\m=a\tilde j^\m$, and we obtain
\disn{di7}{
S_3=-\int\! d^4 x\, \sqrt{-g}\ls\frac{1}{2}\tilde j^\m \tilde j^\n g_{\m\n}+\frac{1}{2}f(a^2)-a\tilde j^\m \ls\dd_\m\la+\al\dd_\m\be\rs\rs.
\nom}
In contrast to the analogs considered in Sections~\ref{rg-prosto},\ref{rg-vihr}, this action is not quadratic in the general case in the scalar variable $a$, but it is still quadratic in the vector variable $\tilde j^\m$.
Therefore we can pass to the analog of the actions \eqref{p9},\eqref{vi9} only in the classical sense (see the text before \eqref{p9}),
i.~e. substituting in \eqref{di7} the value $a$ expressed from the equations of motion. It results in
\disn{di9}{
S_4=-\int\! d^4 x\, \sqrt{-g}\ls\frac{1}{2}\tilde j^\m \tilde j^\n g_{\m\n}-
h(\chi)\rs,\qquad
\chi\equiv\tilde j^\m \ls\dd_\m\la+\al\dd_\m\be\rs,
\nom}
where $h(\chi)$ is a function implicit in the general case.
For the equation of state $p(\ep)=w\ep$, when \eqref{d12} is satisfied, it is easy to calculate:
\disn{d13}{
h(\chi)=\tilde C \chi^{2\frac{1+w}{1+3w}}.
\nom}
For $w=0$ we return to the theory \eqref{vi9} without pressure
(the constant $\tilde C$ can be excluded by rescaling the fields $\la,\al,\be$).
Note that for the action \eqref{di9}, besides the dark energy case $w=-1$, the case $w=-1/3$ also turns out to be special.

Consider now the transition from \eqref{di7} to an analog of the actions \eqref{p12},\eqref{vi12}.
Since \eqref{di7} is quadratic in $\tilde j^\m$, this transition can be understood not only in the classical sense
(as the expressing $\tilde j^\m$ from the equations of motion and the substituting it in the action)
but also in the quantum sense (as a functional integration over $\tilde j^\m$, see the text before \eqref{p9}). As a result we obtain
\disn{di12}{
S_5=-\frac{1}{2}\int\! d^4 x\, \sqrt{-g}\,\Bigl(f(q)-q\, g^{\m\n}\ls\dd_\m\la+\al\dd_\m\be\rs\ls\dd_\n\la+\al\dd_\n\be\rs\Bigr),
\nom}
where the value $a^2$ was replaced by $q$.
In contrast to \eqref{p12},\eqref{vi12}, there is a possibility to further simplify the action \eqref{di12} by excluding the variable $q$.
To do this, we write down the equation obtained by varying \eqref{di12} with respect to $q$:
\disn{d14}{
f'(q)=g^{\m\n}\ls\dd_\m\la+\al\dd_\m\be\rs\ls\dd_\n\la+\al\dd_\n\be\rs,
\nom}
then express $q$ from it and substitute the result in \eqref{di12}. Denoting
\disn{d15}{
\ps\equiv g^{\m\n}\ls\dd_\m\la+\al\dd_\m\be\rs\ls\dd_\n\la+\al\dd_\n\be\rs,
\nom}
we will write the result of the solution of the equation \eqref{d14} with respect to $q$ over $q(\ps)$.
Substituting $q(z)$ in \eqref{di12} and using \eqref{d14} and then \eqref{d9},\eqref{d10},
we obtain the action
\disn{di13}{
\tilde S_5=-\frac{1}{2}\int\! d^4 x\, \sqrt{-g}\,\Bigl(f(q(\ps))-q(\ps)f'(q(\ps))\Bigr)=
\int\! d^4 x\, \sqrt{-g}\,p(\ps),
\nom}
where the pressure $p$ must be expressed in terms of $\ps$.
This action was proposed in \cite{schutz1970}, in which the entropy was also taken into account.
For the given equation of state $p(\ep)$ the inverse dependence $\ps(p)$ can be found using \eqref{d14} and \eqref{d10}:
\disn{d16}{
\ps(p)=f'(q)=\frac{\ep(p)+p}{q(\xi(p))},
\nom}
where $q(\xi)$ is given by \eqref{d11}, and $\xi\equiv \ep(p)-p$.
Inverting this dependence, we found the dependence of the pressure on $\ps$ entering in the action \eqref{di13} for any given equation of state $p(\ep)$.
For the case $p(\ep)=w\ep$ one can show that
\disn{d17}{
p(\ps)=C\ps^\frac{1+w}{2w},
\nom}
where $C$ is the dimensional integration constant determining again the energy of one particle.
For the action \eqref{di13}, besides the dark energy case $w=-1$, the case with no pressure $w=0$ also turns out to be special.

Finally, we write down the generalization of the actions \eqref{p14},\eqref{vi14} as
\disn{di14}{
S_6=-\int\! d^4 x\, \sqrt{-g}\,\Big(\ep(n)-j^\m \ls\dd_\m\la+\al\dd_\m\be\rs+\ga\ls j^\m j^\n g_{\m\n}-n^2\rs\Bigr).
\nom}
If we exclude from it the field $n$ expressed from the obtained motion equation $\ep'(n)=2n\ga$ and substitute the result into the action, then, taking into account \eqref{d9.1} and \eqref{d9}, we obtain the action \eqref{di6}.
If we remove from \eqref{di14} the term containing the fields $\al$ and $\be$ (which ensure the possibility of vortical fluid motion),
then the action will coincide with the expression proposed in \cite{dieztejedor13} (where the entropy was also taken into account).

One can verify that for the actions \eqref{di7},\eqref{di9},\eqref{di12},\eqref{di13},\eqref{di14} the equations of motion result in \eqref{d3} and the EMT results in \eqref{d4}.
Consider this in more detail only for the action \eqref{di13} which has no analogs in the Sections~\ref{rg-prosto},\ref{rg-vihr}. The motion equations for \eqref{di13} are
\disn{d18}{
D_\m \Bigl( 2p'(\ps) g^{\m\n}\ls\dd_\n\la+\al\dd_\n\be\rs\Bigr)=0,\no
\ls\dd_\n\la+\al\dd_\n\be\rs g^{\m\n}\dd_\m\al=0,\qquad
\ls\dd_\n\la+\al\dd_\n\be\rs g^{\m\n}\dd_\m\be=0.
\nom}
It follows from \eqref{d16} that
\disn{d19}{
p'(\ps)=\ls\frac{d\ps}{dp}\rs^{-1}=\ls\frac{\ep'(p)+1}{q}-\frac{(\ep(p)+p)}{q^2}\frac{q}{(\ep(p)+p)}(\ep'(p)-1)\rs^{-1}=\frac{q}{2},
\nom}
where we used the equation \eqref{d11} and the definition of $\xi$. Assume the direction of the fluid motion to be such that
\disn{d20}{
u_\m=\frac{1}{\sqrt{\ps}}\ls\dd_\n\la+\al\dd_\n\be\rs.
\nom}
It follows from \eqref{d14},\eqref{d15} and \eqref{d8},\eqref{d9.1} that $\sqrt{\ps}=\ep'(n)$ and $q\sqrt{\ps}=n$. Taking it into account, it appears that the motion equations \eqref{d18},\eqref{d20} coincide with the motion equations \eqref{d3}.
The same relations allow us to easily verify that the corresponding \eqref{di13} EMT coincides with \eqref{d4}.

\section{Conclusion}\label{zakl}
The variants of the action obtained in this paper describe, in different variables, the same system: a perfect fluid in General Relativity.
Within each of Sections~\ref{rg-prosto},\ref{rg-vihr},\ref{rg-davl}
the actions are fully equivalent, and the variants from different sections differ in the physical properties of the fluid: the presence or absence of vorticity or pressure.
In all cases we considered the fluid isentropic. We did not address the question of whether the discussed theories describe the fluid flow completely or only for a finite period of time, see \cite{dieztejedor13} for this issue.

Among the considered forms of action
the expressions $S_3$ \eqref{p7},\eqref{vi7},\eqref{di7} and $S_4$ \eqref{p9},\eqref{vi9}, \eqref{di9}
are completely new (as far as the author knows).
The generalizations of the action $S_5$ \eqref{p12} (which was studied in
\cite{Golovnev201439}) on the cases of the presence of vorticity \eqref{vi12}
and pressure \eqref{di12} also have not been considered previously.
Note that in the absence of pressure the actions $S_3$, $S_4$, $S_5$, $S_6$
are polynomial with respect to the variables describing the perfect fluid.
At that the action $S_3$ \eqref{p7},\eqref{vi7} has an especially simple structure: it
contains quadratic terms with respect to variables $\tilde j^\m$ and $a$,
while the remaining terms depend on these variables linearly.

Some of the considered methods of describing a perfect fluid may be more suitable than others in various problems concerning a fluid flow.
For further development of the ideas of mimetic gravity, the actions $S_5$ \eqref{vi12},\eqref{di12}, containing only scalar fields, as well as
the variant of the theory with nonpotentially moving mimetic matter \eqref{vi15},\eqref{vi16} described in Section~\ref{rg-mimet} can be especially useful.

To study the phenomenology of particle creation in gravitational fields \cite{berezin17}, the forms of the action from Section \ref{rg-davl} can be useful, for which the motion is sufficiently arbitrary and the pressure is present.
Adding to them the terms dependent on the scalar field $\la$, we can introduce the creation of particles into the theory, since such an additive leads to a deviation from the exact fulfillment of the continuity equation.

Some of the considered forms of the action can also be useful in developing the idea of describing the gravity within the framework of the embedding theory
\cite{regge,deser,statja18,statja25},
where the space-time is considered as a four-dimensional surface in a flat ambient space (bulk).
As shown in \cite{statja33}, there is a similarity between the approaches of the mimetic gravity and the embedding theory,
because both theories are equivalent to General Relativity in the presence of some additional (really not existing) matter.
The action of embedding theory can be written as a sum of Einstein-Hilbert
term and the term that describes such additional matter:
\disn{za1}{
S=-\frac{1}{2}\int\! d^4 x\, \sqrt{-g}\,
\Bigl( g_{\m\n}-(\dd_\m y^a)(\dd_\n y_a)\Bigr)\tau^{\m\n}.
\nom}
Here  $y^a(x^\m)$ is the embedding function that defines a
four-dimensional surface and $\tau^{\m\n}$ is the Lagrange multiplier,
which turns out to be the EMT of the additional matter.
One can attempt to relate some of the variants of the perfect fluid description considered in this paper
(e.~g. the actions $S_4$ \eqref{p9},\eqref{vi9}, the expressions of which are close to \eqref{za1})
to the additional matter arising in the embedding theory.

{\bf Acknowledgements.}
The author is grateful to
A.~Starodubtsev and A.~Golovnev for useful discussion and to
V.~Berezin for useful references.
The work was supported by Saint Petersburg State University research grant No.~11.38.223.2015.

\providecommand{\eprint}[1]{\href{http://arxiv.org/abs/#1}{\texttt{#1}}}

\end{document}